\documentclass{emulateapj}

\usepackage{epsfig}
\usepackage{amsmath}
\usepackage{rotating}
\usepackage{natbib}
\usepackage{lscape}
\usepackage{enumerate}
\usepackage{multirow}
\usepackage{array}
\usepackage{appendix}

\def\plotone#1{\centering \leavevmode
\includegraphics[clip=, width=.85\columnwidth]{#1}}

\def\plottwo#1#2{\centering \leavevmode
\includegraphics[width=.45\columnwidth]{#1} \hfil
\includegraphics[width=.45\columnwidth]{#2}}

\newcommand{\cN}[1]{\mathcal{N}}

\def\gsim{\;\rlap{\lower 2.5pt
 \hbox{$\sim$}}\raise 1.5pt\hbox{$>$}\;}
\def\lsim{\;\rlap{\lower 2.5pt
   \hbox{$\sim$}}\raise 1.5pt\hbox{$<$}\;}

\begin{document}


\title{Atmosphere and Spectral Models of the {\it Kepler}-Field
  Planets HAT-P-7\lowercase{b} and TrES-2}

\author{
David S. Spiegel\altaffilmark{1,2},
Adam Burrows\altaffilmark{1,2}
}

\affil{$^1$Department of Astrophysical Sciences, Peyton Hall,
  Princeton University, Princeton, NJ 08544}
 \affil{$^2$Kavli Institute for Theoretical Physics, UCSB, Santa
   Barbara, CA 93106-4030}

\vspace{0.5\baselineskip}

\email{
dsp@astro.princeton.edu,
burrows@astro.princeton.edu,
}

\begin{abstract}
We develop atmosphere models of two of the three {\it Kepler}-field
planets that were known prior to the start of the {\it Kepler} mission
(HAT-P-7b and TrES-2).  We find that published {\it Kepler} and {\it
  Spitzer} data for HAT-P-7b appear to require an extremely hot upper
atmosphere on the dayside, with a strong thermal inversion and little
day-night redistribution.  The {\it Spitzer} data for TrES-2 suggest a
mild thermal inversion with moderate day-night redistribution.  We
examine the effect of nonequilibrium chemistry on TrES-2 model
atmospheres and find that methane levels must be adjusted by extreme
amounts in order to cause even mild changes in atmospheric structure
and emergent spectra.  Our best-fit models to the {\it Spitzer} data
for TrES-2 lead us to predict a low secondary eclipse planet-star flux
ratio ($\lsim$$2\times 10^{-5}$) in the {\it Kepler} bandpass, which
is consistent with what very recent observations have found.  Finally,
we consider how the {\it Kepler}-band optical flux from a hot
exoplanet depends on the strength of a possible extra optical absorber
in the upper atmosphere.  We find that the optical flux is not
monotonic in optical opacity, and the non-monotonicity is greater for
brighter, hotter stars.
\end{abstract}


\section{Introduction}
\label{sec:intro}
Extrasolar planets are being discovered at an increasingly rapid pace:
roughly a quarter of the currently known exoplanets (numbering more
than 450, as of June, 2010) were found since the beginning of
2009.\footnote{See the catalogs at http://exoplanet.eu and
  http://www.exoplanets.org.}  Still, only $\sim$80 of the known
planets have been seen to transit their parent stars.  Transits break
the degeneracy between mass and inclination, they allow direct
measurement of planetary radii, and they make possible precise
measurements of planetary fluxes from secondary eclipse observations.
The {\it Kepler} mission, which is predicted to find many new
transiting planets, is, therefore, particularly exciting.

Three transiting planets in the {\it Kepler} field were identified
prior to the beginning of the {\it Kepler} mission -- TrES-2,
HAT-P-7b, and HAT-P-11b (occasionally referred to as Kepler-1b, -2b, and
-3b).  \citet{spiegel_et_al2010a} have already published a range of
possible atmospheric models of HAT-P-11b; here, we consider HAT-P-7b
and and TrES-2.

The InfraRed Array Camera (IRAC) instrument on the {\it Spitzer} Space
Telescope has been a boon to exoplanetary science, providing
observations that are diagnostic of atmospheric temperature and
composition for more than a dozen planets.  It had four photometric
channels, centered at 3.6~$\mu$m, 4.5~$\mu$m, 5.8~$\mu$m, and
8.0~$\mu$m.  Recently, \citet{christiansen_et_al2010} and
\citet{odonovan_et_al2010} used IRAC to measure infrared fluxes from
HAT-P-7b and TrES-2, respectively.

HAT-P-7b, discovered by \citet{Pal_et_al_2008}, orbits a large, hot
star ($1.84R_\sun$, spectral type F8).  It is one of the most highly
irradiated known explanets, with a substellar flux of $\sim$$4.8\times
10^9\rm~erg~cm^{-2}~s^{-1}$.  Its orbit is significantly misaligned
from the stellar spin vector, indicating a possible third body in the
system \citep{winn_et_al2009, Narita_et_al_2009_2}.  It is a
particularly interesting object in part because {\it Kepler's}
exquisite photometry has allowed measurement of ellipsoidal variations
in the star induced by the planet's tidal field
\citep{welsh_et_al2010}.  TrES-2, by contrast, orbits a nearly
solar-type star in a nearly grazing orbit \citep{odonovan_et_al2006,
  holman_et_al2007, sozzetti_et_al2007, raetz_et_al2009}.
\citet{mislis+schmitt2009} find a reduction in transit duration of
$\sim$3 minutes since 2006, and attribute this shortening to a change
in inclination, although analysis by \citet{rabus_et_al2009} does not
corroborate such a large change.

The atmosphere modeling strategy that we employ here differs from
several others that have been used.  Similar to
\citet{barman_et_al2005} and \citet{fortney_et_al2006}, we calculate
radiative equilibrium, chemical equlibrium models.  In contrast, both
\citet{madhusudhan+seager2009} and \citet{tinetti_et_al2005,
  tinetti_et_al2007} adjust both chemistry and thermal structure in
order to find a best fit to the available data, eschewing equilibrium
solutions.  This latter method produces chemical and thermal profiles
that are not the result of ab initio calculations, but that might
reveal non-equilibrium behavior.  Yet another approach is taken by
\citet{showman_et_al2009} and \citet{burrows_et_al2010}, who simulate
three dimensional structure and dynamics in planetary atmospheres; so
far this more sophisticated approach has not produced better fits to
observations than the one-dimensional radiative models described
above.

\citet{hubeny_et_al2003} were the first to suggest that an extra
optical absorber in a hot exoplanet's upper atmosphere could lead to a
thermal inversion.  Observations seem to suggest such inversions
\citep{hubeny_et_al2003, fortney_et_al2006, fortney_et_al2008,
  burrows_et_al2007c, burrows_et_al2008b, richardson_et_al2007,
  spiegel_et_al2009b, madhusudhan+seager2009, knutson_et_al2008b,
  knutson_et_al2010}, and the strong optical absorber titanium oxide
(TiO) has frequently been suggested as the possible culprit
responsible for the inferred inversions.  However, in the absence of
strong mixing processes, the molecular weight of TiO would tend to
make it settle to the bottom of the atmosphere.  Furthermore, cold
traps on the nightside and below the hot upper atmosphere on the
dayside can cause TiO to condense into solid grains, which
necessitates even stronger macroscopic mixing to keep TiO aloft in the
radiatively important upper atmosphere.  Since the photospheres are
above the radiative-convective boundaries, they are stably stratified;
it is not obvious whether such strong mixing obtains in such a stable
region \citep{spiegel_et_al2009b}.  Various authors have tried to
estimate the amount of macroscopic mixing and have found that, in some
regions of the day-side atmosphere, the mixing might be vigorous
enough to maintain TiO at altitude \citep{showman_et_al2009,
  li+goodman2010, madhusudhan+seager2010}, though
\citet{youdin+mitchell2010} point out that turbulent diffusivity in
excess of $10^7 \rm~cm^2~s^{-1}$ might lead to overinflation of some
planets through downward transport of entropy.
\citet{zahnle_et_al2009} suggest that sulfur photochemistry provides
another avenue for achieving the additional upper-atmosphere optical
opacity that is needed to produce hot upper atmospheres and
inversions.  \citet{knutson_et_al2010} find that the presence of
thermal inversions appears to be inversely related to the host stars'
ultraviolet (UV) activity, raising the possibility that strong
incident UV destroys molecules that may be responsible for thermal
inversions.  Here, we remain agnostic on the matter and include, for
modeling purposes, an extra source of optical opacity of as-yet
unknown origin.

This paper is structured as follows.  In \S\ref{sec:methods}, we
describe our 1D atmosphere modeling strategy.  In \S\ref{sec:H7} and
\S\ref{sec:T2}, we present our models of HAT-P-7b and TrES-2.  We
point out the effects that scattering and nonequilibrium chemistry
could have on our models.  In \S\ref{sec:nonmon}, we discuss the
non-monotonic relationship between the optical opacity in a planet's
upper atmosphere and the planet's optical emergent radiation.
Finally, in \S\ref{sec:conc}, we conclude, and in an Appendix, we
discuss the various methods used in the literature to represent
day-night redistribution in this modeling context.

\section{Atmosphere Modeling}
\label{sec:methods}
As in our other recent studies, we use the code {\tt COOLTLUSTY}
\citep{hubeny_et_al2003, sudarsky_et_al2003, burrows_et_al2006,
  burrows_et_al2008b, spiegel_et_al2009b, spiegel_et_al2010a}, a
variant of the code {\tt TLUSTY} \citep{hubeny1988, hubeny+lanz1995},
to calculate radiative equilibrium irradiated atmosphere models.  Our
atomic and molecular opacities are generally calculated assuming
chemical equilibrium with solar elemental abundances
\citep{sharp+burrows2007, burrows+sharp1999, burrows_et_al2001,
  burrows_et_al2002, burrows_et_al2005}, although we also calculate a
few nonequilibrium models, described in \S\ref{sec:T2}.  The
irradiating spectra in our models are taken from \citet{kurucz1979,
  kurucz1994, kurucz2005}, interpolated to the effective temperatures
and surface gravities of HAT-P-7 and the host star of TrES-2.

\begin{figure}
\plottwo
{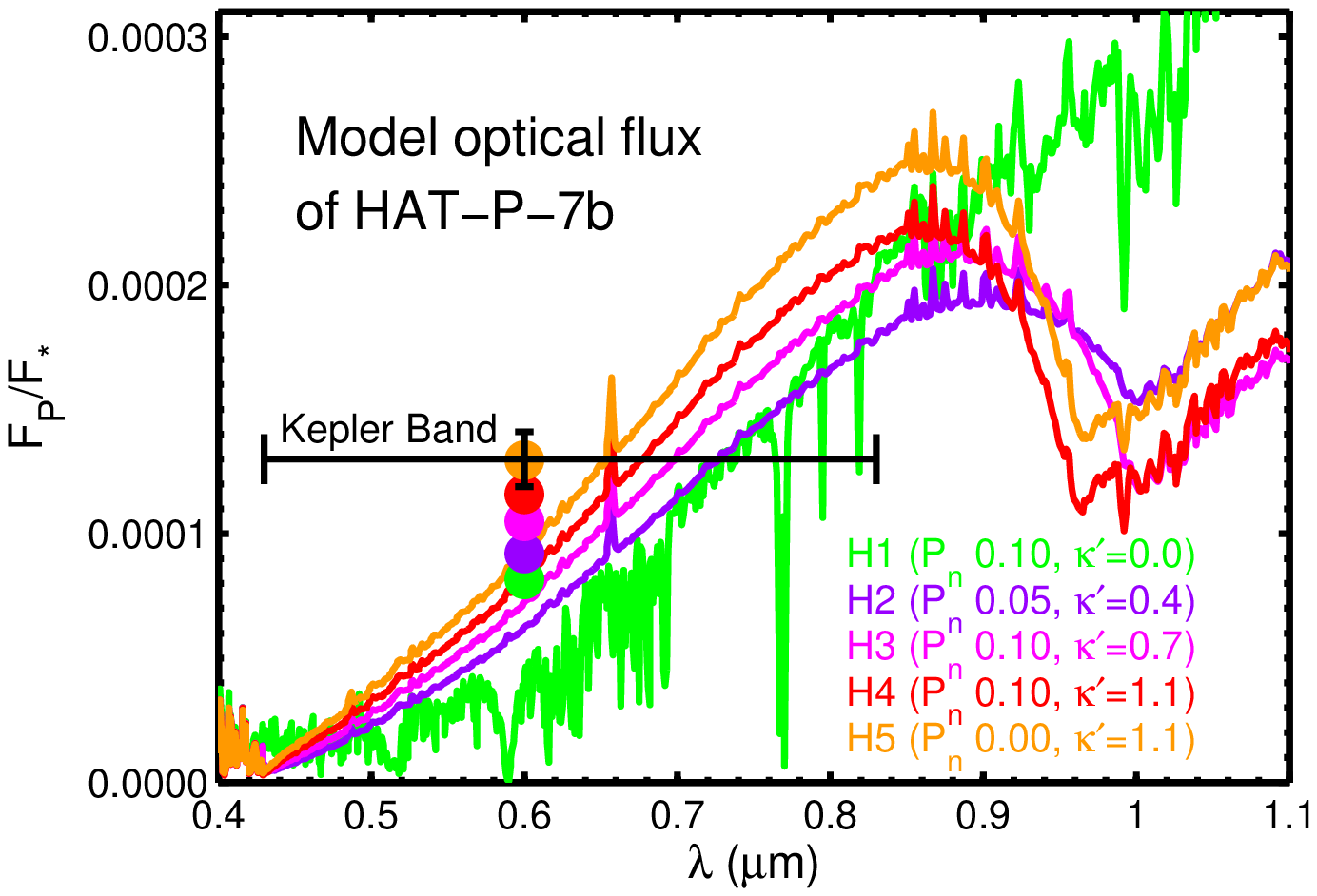}
{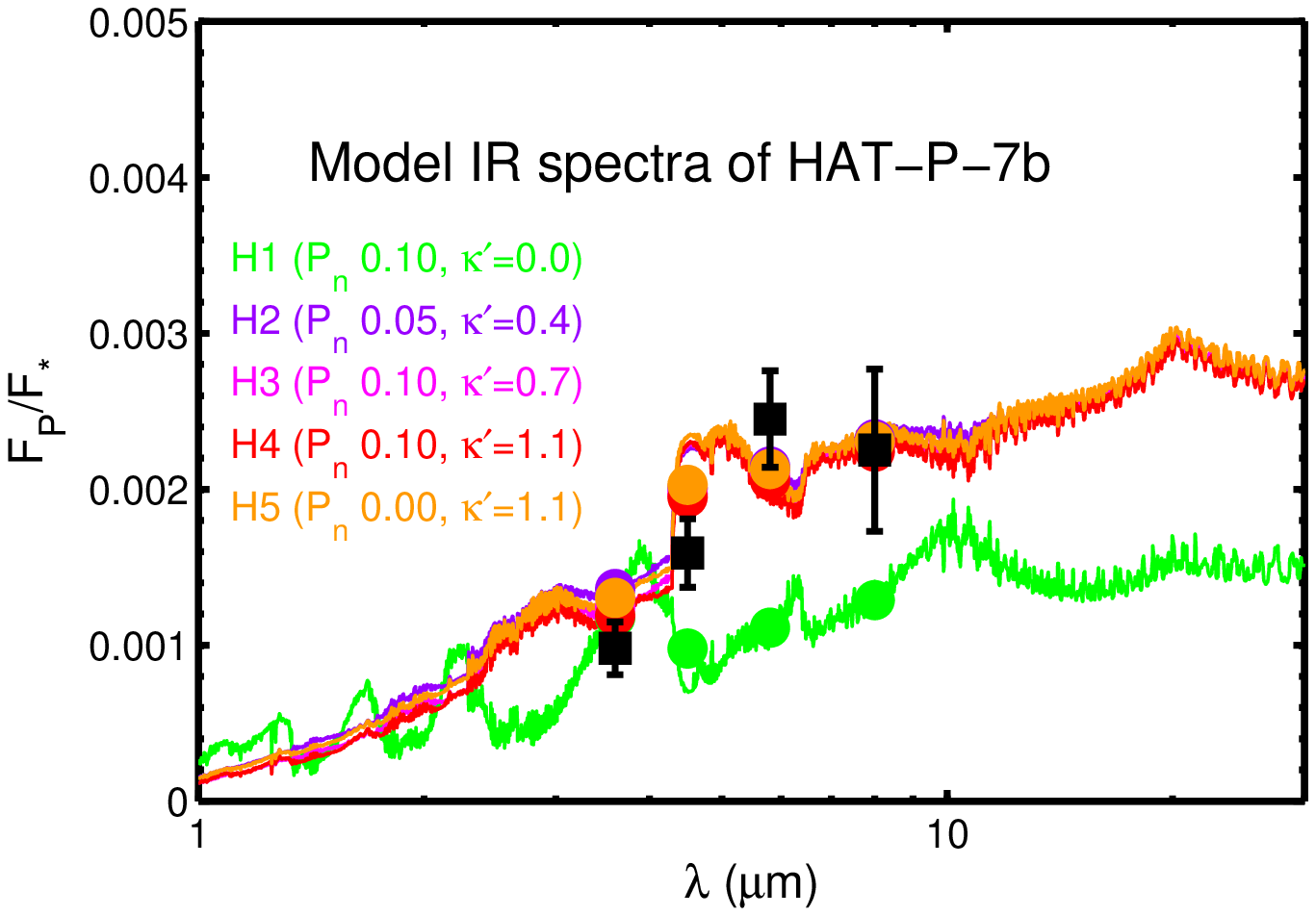}
\plotone
{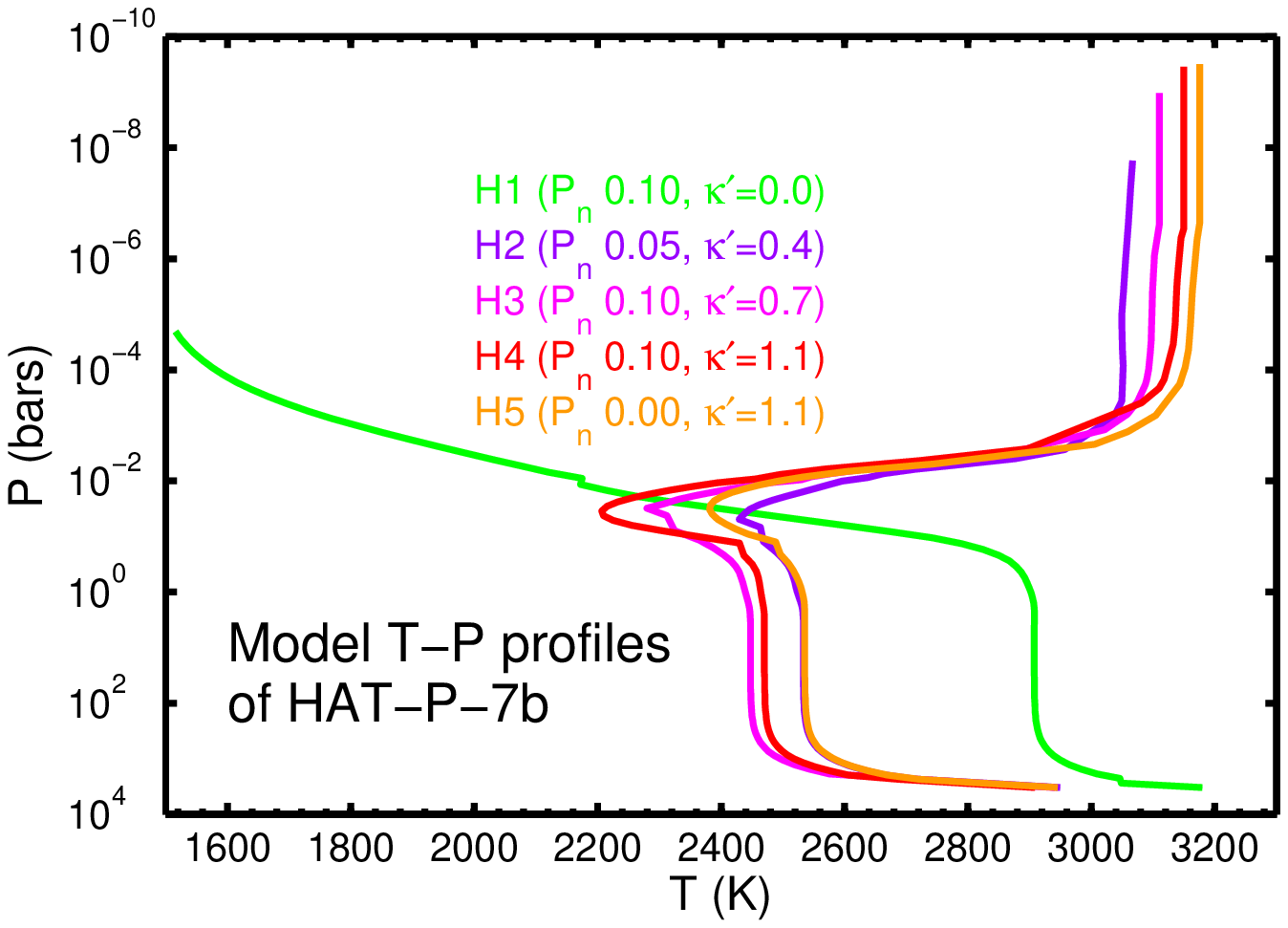}
\caption{HAT-P-7b model atmospheres.  The models (H1--H5) encompass
  varying degrees of redistribution (parameterized as $P_n$) and
  upper-atmosphere optical absorption (parameterized as $\kappa'$,
  where the values are in cm$^2$~g$^{-1}$), and are listed in
  Table~\ref{ta:models}.  Measured data are shown where available.
  {\it Top left:} Optical planet-star flux ratios.  The data point
  (indicated by the 1-$\sigma$ error bars) is from
  \citet{borucki_et_al2009}, and the horizontal black line indicates
  the full width at 10\% maximum of the {\it Kepler} bandpass.  The
  filled colored circles show the integrals of the model flux over the
  {\it Kepler} bandpass.  The model that fits the data best is H5,
  which has significant upper-atmosphere absorption ($\kappa' = 1.1
  {\rm~cm^2~g^{-1}}$) and no redistribution.  {\it Top Right:}
  Infrared planet-star flux ratios.  The data points with 1-$\sigma$
  error bars are IRAC data from \citet{christiansen_et_al2010}.
  Filled colored circles show the integrals of models over the
  response functions for the IRAC bands.  Among models with
  significant upper-atmosphere absorption (models H2--H5), there is
  little difference in the infrared spectrum.  Model H5 is a slightly
  better fit than H2, a slightly worse fit than H3 and H4 at
  3.6~$\mu$m, and is a slightly worse fit than models H2--H4 at
  4.5~$\mu$m.  The models fit the 5.8-$\mu$m and 8.0-$\mu$m IRAC data
  equally well.  {\it Bottom:} Temperature-pressure profiles.  The
  models with extra optical absorption in the upper atmosphere (models
  H2--H5) all show large thermal inversions, with upper-atmosphere
  temperatures $\sim$1500~K or more hotter than what would be expected
  in the absence of extra heating (indicated by model H1).  Note that
  our models with an extra absorber extend to very low pressures
  ($\sim$10$^{-8}$~bars), where other physics in the thermosphere
  might be important in determining the actual radiative equilibrium
  thermal profile.}
\label{fig:H7}
\end{figure}

In addition to the observationally measured parameters of our models
(orbital semimajor axis, planet and stellar radii, planet and stellar
surface gravities, stellar effective temperature), and in addition to
calculating equilibrium chemistry and radiative transfer, several
other physical processes go into our models.  These include Rayleigh
scattering \citep{sudarsky_et_al2000, lopez-morales+seager2007,
  burrows_et_al2008, Rowe_et_al_2008}, heat redistribution, and the
possible presence of an extra optical absorber that could explain the
hot upper atmospheres and thermal inversions that have been inferred
from infrared observations of several transiting planets.  In
particular, there are two key free parameters that we vary: $P_n$ and
$\kappa'$.
\vspace{-0.10in}
\begin{itemize}
\item $P_n$ quantifies the efficiency of day-to-night heat
  redistribution, and is equal to the fraction of incident day-side
  heating that is reradiated from the nightside
  \citep{burrows_et_al2006, burrows_et_al2008b}.  In the models in
  this paper, the redistribution takes place between 0.01 and 0.1
  bars.
\item $\kappa'$ is an ad hoc extra source of optical absorption
  opacity in the upper atmosphere, and is motivated by the thermal
  inversions that have been inferred from the infrared spectra of many
  exoplanets.  $\kappa'$ is similar to the $\kappa_e$ of some of our
  recent work (e.g., \citealt{burrows_et_al2008b}), except that,
  rather than a gray optical absorber, it has the same parabolic
  dependence on frequency as the corresponding extra absorber in
  \citet{lopez-morales_et_al2009} and \citet{burrows_et_al2010}.
 \end{itemize}
The models whose properties are plotted in
Figs.~\ref{fig:H7}--\ref{fig:T2n} are summarized in
Table~\ref{ta:models}.  In addition to these, we also calculate
several models with a modification to the code that allows an ad hoc
extra source of optical scattering opacity, analogous to $\kappa'$,
but for scattering instead of absorption.

\section{HAT-P-7\lowercase{b}}
\label{sec:H7}
HAT-P-7b is a $1.78M_J$, $1.36R_J$ planet in an approximately circular
orbit 0.0377~AU from the 1.47~$M_\sun$, 1.84~$R_\sun$, F6 star HAT-P-7
\citep{Pal_et_al_2008}.  In addition to being in the {\it Kepler}
field, HAT-P-7b is interesting because Rossiter-McLaughlin
measurements indicate that it is in a polar or retrograde orbit
\citep{winn_et_al2009, Narita_et_al_2009_2}.  More relevant to
modeling its atmosphere is the fact that, due to its close proximity
to a large, relatively hot ($\sim$6350~K) star, HAT-P-7b experiences
unusually high stellar irradiation ($\sim$$4.8\times
10^9\rm~erg~cm^{-2}~s^{-1}$ at the substellar point).

As an early confirmation that {\it Kepler} was performing well,
\citet{borucki_et_al2009} published 10 days of commisioning-phase data
on HAT-P-7b.  These data reveal a surprisingly large secondary eclipse
depth in the {\it Kepler} band ($\sim$0.43-0.83~$\mu$m), as the
corresponding planet-star flux ratio decreases by $(1.3\pm 0.1)\times
10^{-4}$ when the planet passes behind the star.\footnote{This
  contrast ratio is similar to the expected dip in flux for an
  Earth-sized planet passing in front of a Sun-like star, and
  therefore indicated that {\it Kepler} should be capable of
  performing the mission for which it was designed.}
\citet{borucki_et_al2009} suggest that such a large contrast ratio
could imply that the atmosphere absorbs strongly and has minimal
redistribution to the night side (low $P_n$, in our language).  They
estimate a day-side temperature of $2650\pm 100$~K.

Infrared data for HAT-P-7b became available shortly thereafter, when
\citet{christiansen_et_al2010} presented secondary eclipse
observations of the planet employing the IRAC instrument on the {\it
  Spitzer} Space Telescope.  They found planet-star flux-ratios of
$(9.8\pm 1.7)\times 10^{-4}$, $(15.9\pm 0.22)\times 10^{-4}$,
$(24.5\pm 3.1)\times 10^{-4}$, and $(22.5\pm 5.2)\times 10^{-4}$ at
the IRAC 3.6-$\mu$m, 4.5-$\mu$m, 5.8-$\mu$m, and 8.0-$\mu$m, channels,
respectively.

Although, as \citet{christiansen_et_al2010} argue, HAT-P-7b's
atmosphere is irradiated strongly enough that it is probably too hot
for the condensates that might otherwise be expected to contribute
significant scattering opacity, we nevertheless tried adding an ad hoc
extra scatterer to the upper atmosphere of some models.  We find that
the optical point can easily be matched by a model with significant
optical scattering, but such models drastically underpredict the {\it
  Spitzer} data.  Since the \citet{borucki_et_al2009} speculation of
an extremely hot upper atmosphere with low albedo came before the
infrared data were published, it might have been a little bit
premature.  On the other hand, the expectation of low albedo is well
motivated by the high stellar irradiation, and seems to be confirmed
by the infrared observations.

In Table~\ref{ta:models}, we present five thermochemical equilibrium
models of HAT-P-7b, four having extremely hot upper atmospheres
(ranging from H2's $\sim$3040~K to H5's $\sim$3180~K), and a
comparison model without a thermal inversion (H1).  These five models
span a (small) range of values of both $P_n$ (the degree of day-night
redistribution) and $\kappa'$ (the strength of an ad hoc extra
absorber, where the values are in cm$^2$~g$^{-1}$).  Models with
values of $P_n$ larger than 0.1 are not displayed in this table
because such models underpredict the optical data.

The five HAT-P-7b models of Table~\ref{ta:models} are displayed in
Fig.~\ref{fig:H7}.  The top-left and top-right panels portray the
wavelength-dependent planet-star flux ratios in the optical and the
infrared, respectively, and the corresponding data
\citep{borucki_et_al2009, christiansen_et_al2010} are superimposed.
In both panels, planet and star fluxes are integrated over the
relevant bandpasses, and the resulting integrated planet-star ratios
are displayed as solid filled circles.  The bottom panel shows
temperature-pressure profiles for these five models.  Models H2-H5,
with extra optical opacity, have strong thermal inversions in which
the upper atmosphere is heated to temperatures $\gsim$1500~K greater
than they would be in the absence of the extra absorber (as
represented by model H1).\footnote{The temperature-pressure profiles
  for models H2-H5 are shown up to extremely low pressures
  ($\sim$10$^{-8}$~bars), at which point nonequilibrium processes,
  such as photochemical dissociation and UV opacity, would be
  important for determining the true thermal profiles.}  The near- and
mid-infrared spectra for inverted models H2-H5 are nearly
indistinguishable from one another, and are all reasonable fits to the
IRAC data (although the models are $\sim$1-2$\sigma$ higher than the
data at 3.6~$\mu$m and 4.5~$\mu$m).  Taking into account the optical
data (top-left panel), model H5 is the best fit to the available data.
HAT-P-7b's real atmosphere is probably not represented by a 1D
radiative equilibrium model, such as model H5, but the available data
suggest that significant flux might be absorbed high in the
atmosphere, where the radiative timescale is short
\citep{iro_et_al2005, showman_et_al2008}, which would imply that
day-side heat would probably be reradiated before much advective
redistribution has occurred.  The association of high extra opacity
with low day-night redistribution in the best-fitting model,
therefore, is consistent with what should be expected.

\citet{christiansen_et_al2010} analyzed the thermochemical
implications of the {\it Kepler}/{\it Spitzer} data, as well.  Using
the method described in \citet{madhusudhan+seager2009} and the four
IRAC data points in conjunction with the {\it Kepler} data, they find
classes of models that fit the five data points optimally.  They find
that a blackbody temperature of $\sim$3175~K is needed to fit the {\it
  Kepler} data, significantly higher than the brightness temperature
inferred by \citet{borucki_et_al2009}; lower brightness temperatures
are inferred for the IRAC data.  Their best-fit (non-blackbody) models
are (by construction) not in local radiative or chemical equilibrium,
but all of the best-fitting models contain thermal inversions, as do
our models H2--H5.  By relaxing the fit-criterion to 1.25-$\sigma$ at
each datum and allowing strongly nonequilibrium chemical abundances,
they find a noninverted model with significant CH$_4$ abundance, but
the high temperatures of the atmosphere favor CO and therefore favor
the inverted models, consistent with our analysis.

\section{TrES-2}
\label{sec:T2}
TrES-2 is a $1.20M_J$, $1.22R_J$ planet in nearly circular orbit
0.0356~AU from its 1.06~$M_\sun$, 1.00~$R_\sun$ solar-type (G0V) star
\citep{odonovan_et_al2006}.  At its substellar point, TrES-2
experiences an irradiating flux of $\sim$1.1$\times
10^9\rm~erg~cm^{-2}~s^{-1}$, similar to that experienced by
HD~209458b.

\citet{odonovan_et_al2010} report IRAC observations of TrES-2,
together with nonequilibrium atmosphere models of the planet generated
in the manner of \citet{madhusudhan+seager2009}.
\citeauthor{odonovan_et_al2010} find that the infrared data can be
reasonably well fit by a blackbody model, a model with a thermal
inversion, and a model without a thermal inversion.  They point out
that their model without a thermal inversion requires a surprisingly
low abundance of CO, given the temperature of the atmosphere
($\sim$1500~K).  Therefore, they favor the model with the inversion.

Here, we consider a variety of models of TrES-2.  Motivated by the
analysis of \citet{odonovan_et_al2010}, we examine models both with
equilibrium and with nonequilibrium chemistry, all of which are in
radiative equilibrium.  In Table~\ref{ta:models}, we list five TrES-2
models with opacities defined by equilibrium chemistry at solar
abundances (T1--T5) and five that are completely analogous, except
with the CO abundance artificially set to 0 (T1n--T5n).  In the latter
group of models, the carbon that would have been in CO is instead in
CH$_4$, and the excess oxygen is instead in H$_2$O.  Among both the
set of models with and without CO, there is a model that has no extra
absorber and no inversion (T1 and T1n, respectively), and four models
that do have an extra absorber and thermal inversions.  After we had
generated these atmosphere models, \citet{croll_et_al2010} and
\citet{kipping+bakos2010} published {\it Ks}-band ($\sim$2.2~$\mu$m)
and {\it Kepler}-band observations, respectively, of TrES-2's
secondary eclipse.  The models presented herein can, therefore, be
thought of as predictions, not ``postdictions,'' for the recent data.
As a result, it was gratifying to see that the new data are consistent
with our predictions, as described below.

\begin{figure}
\plottwo
{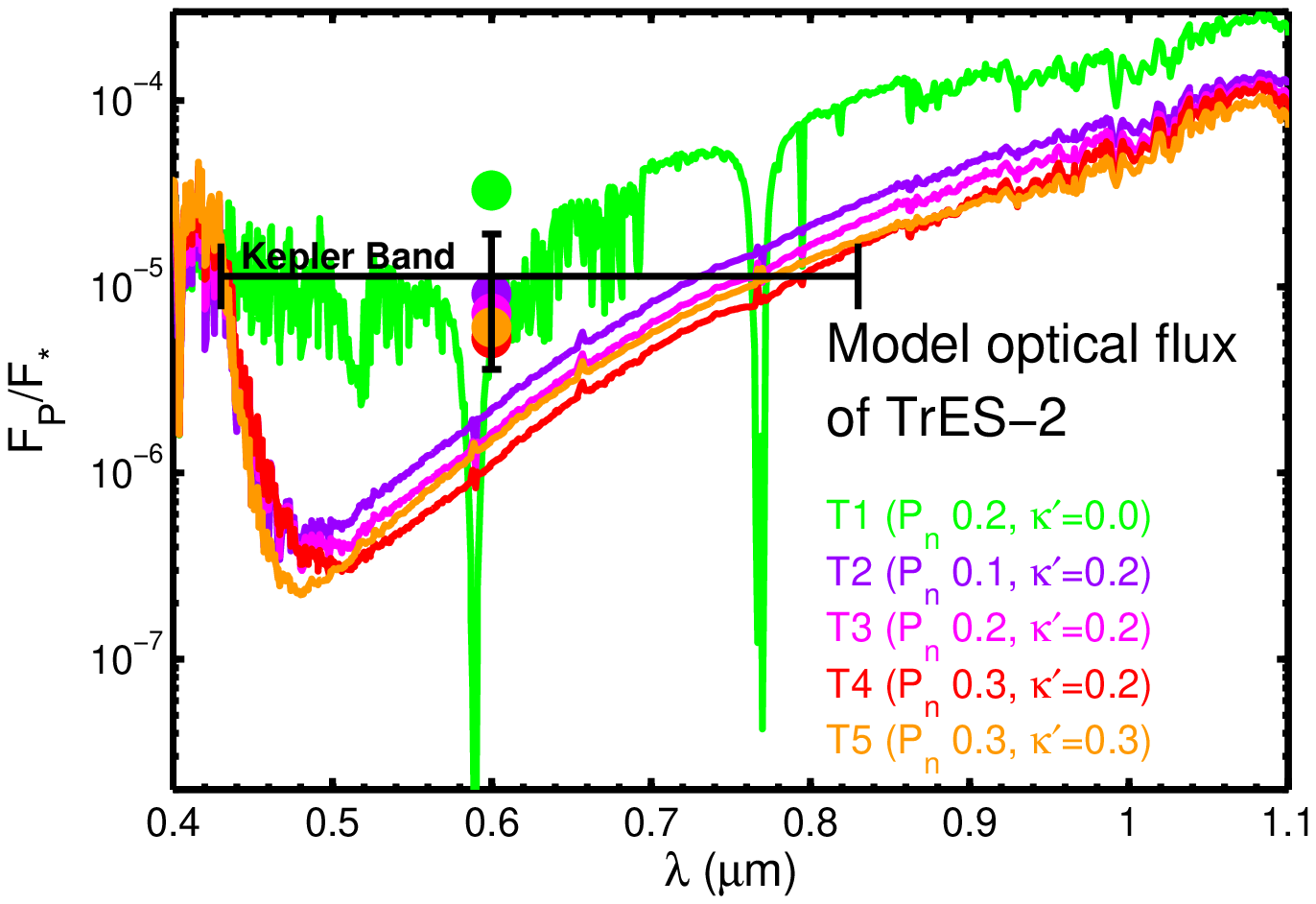}
{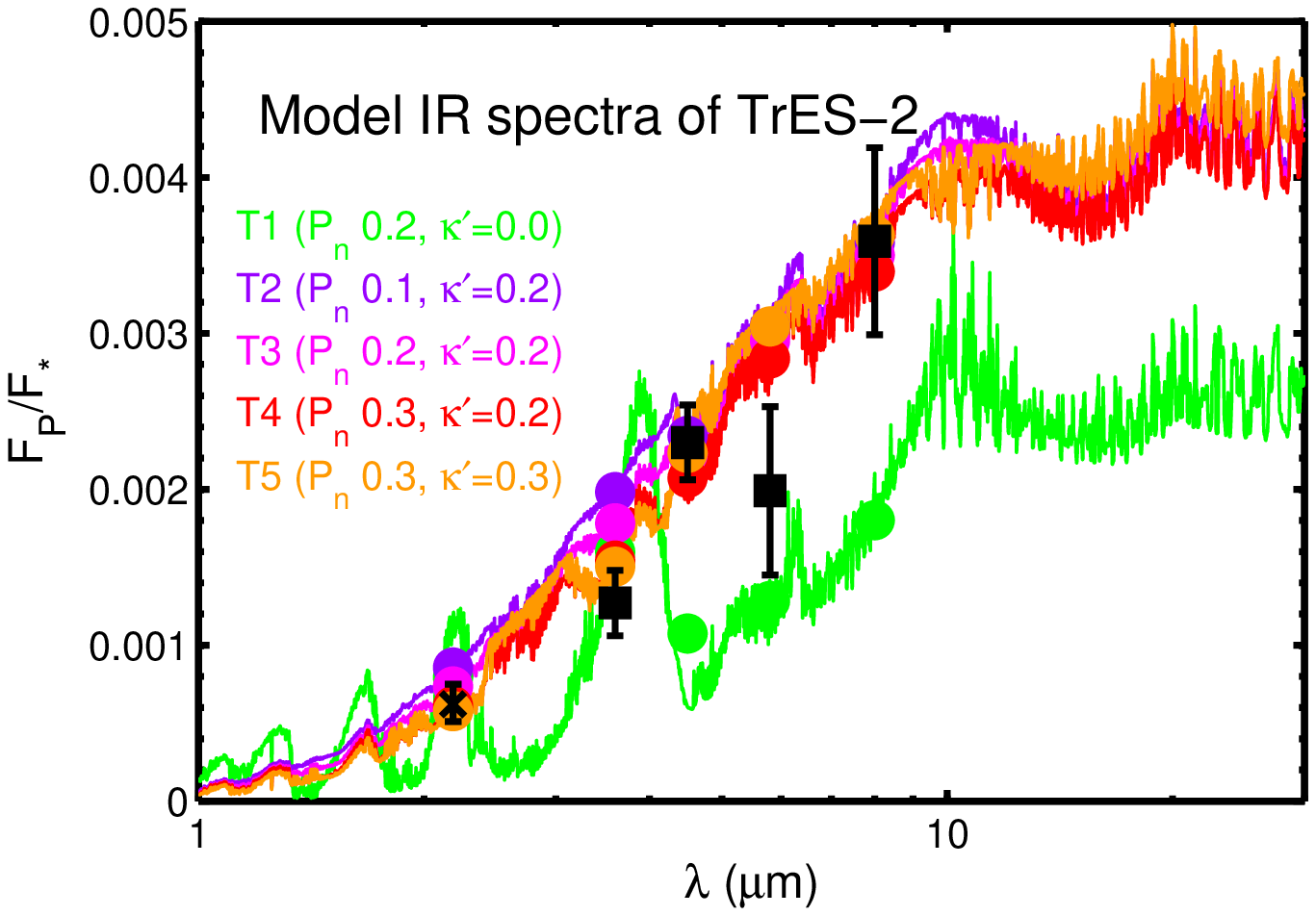}
\plotone
{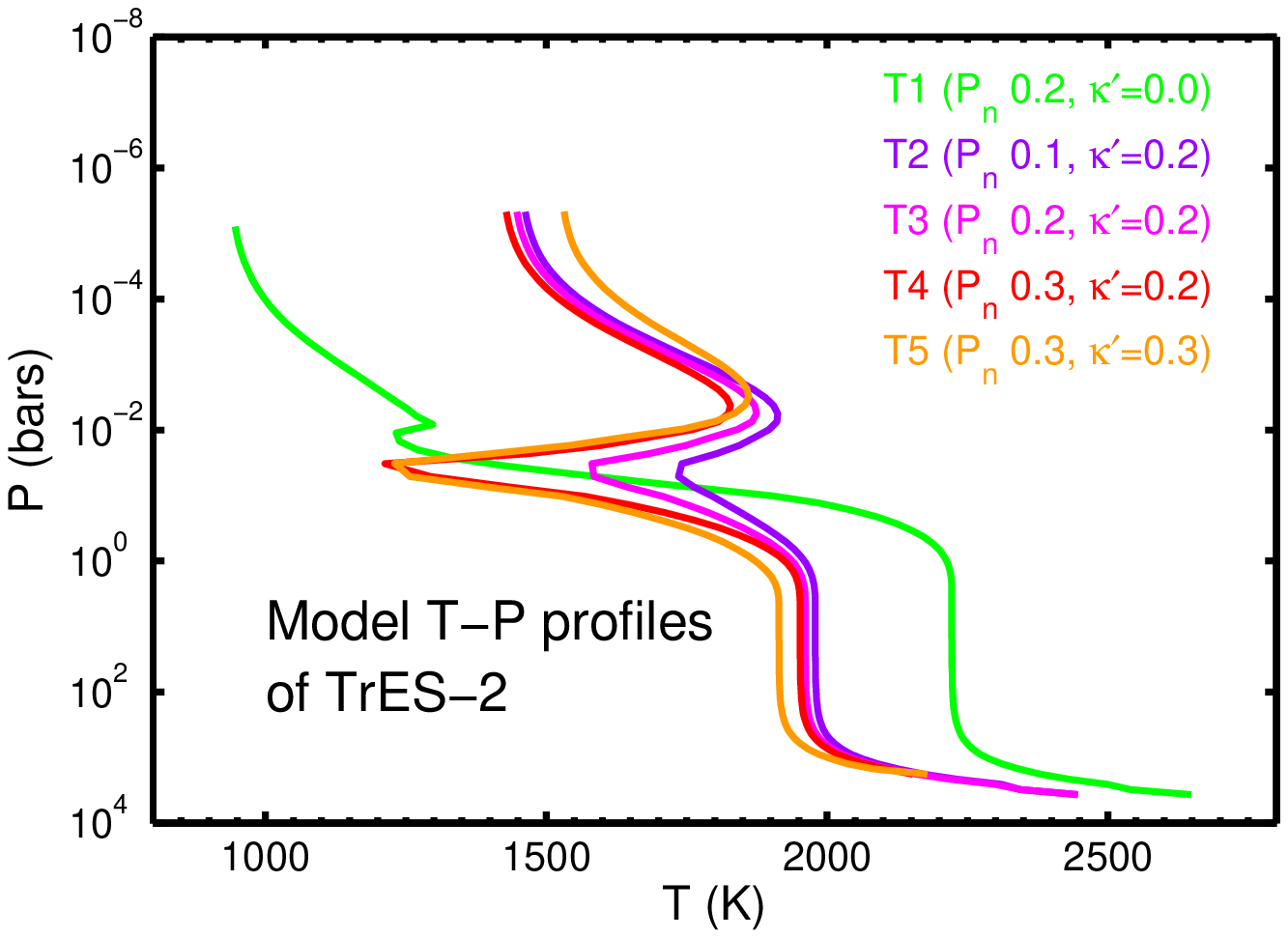}
\caption{TrES-2 equilibrium chemistry model atmospheres, analogous to
  Fig.~\ref{fig:H7}, but for TrES-2 instead of HAT-P-7b.  The models
  (T1--T5) encompass varying degrees of heat redistribution ($P_n$)
  and upper-atmosphere optical absorption ($\kappa'$, values in
  cm$^2$~g$^{-1}$), and are listed in Table~\ref{ta:models}.  {\it Top
    left:} Optical planet-star flux ratios (log scale).  The
  horizontal black line indicates the full width at 10\% maximum of
  the {\it Kepler} bandpass, and is at the level of the data (shown
  with with 2-$\sigma$ vertical error bars) from
  \citet{kipping+bakos2010}.  The filled colored circles show the
  integrals of the model flux over the {\it Kepler} bandpass.  Models
  with upper atmosphere absorption (T2--T5) tend to have low optical
  flux ($F_p/F_* \lsim 2\times 10^{-5}$).  Even with $\kappa' =
  0\rm~cm^2~g^{-1}$ (model T1), the optical flux ratio is
  $\sim$3$\times 10^{-5}$.  Models T2--T5 are all consistent with the
  data, to 2~$\sigma$.  {\it Top Right:} Infrared planet-star flux
  ratios.  The black square data points with 1-$\sigma$ error bars are
  IRAC data from \citet{odonovan_et_al2010}, and the black cross data
  point (with 1-$\sigma$ error bars) represents {\it Ks}-band data
  from \citet{croll_et_al2010}.  Filled colored circles show the
  integrals of models over the response functions for the IRAC bands
  (at 3.6, 4.5, 5.8, and 8.0~$\mu$m) and for {\it Ks}-band (at
  2.2~$\mu$m).  The 4.5-$\mu$m, 5.8-$\mu$m, and 8.0-$\mu$m IRAC data
  clearly favor models with some extra absorption.  At 3.6~$\mu$m,
  4.5~$\mu$m, and 8.0~$\mu$m, model T5 is a good fit to the data, but
  it misses the data by $\sim$2$\sigma$ at 5.8~$\mu$m.  {\it Bottom:}
  Temperature-pressure profiles.  The models with extra optical
  absorption in the upper atmosphere (models T2--T5) show moderate
  thermal inversions, with upper-atmosphere temperatures $\sim$500~K
  hotter than what would be expected in the absence of extra heating
  (indicated by model T1).}
\label{fig:T2}
\end{figure}

\begin{figure}
\plottwo
{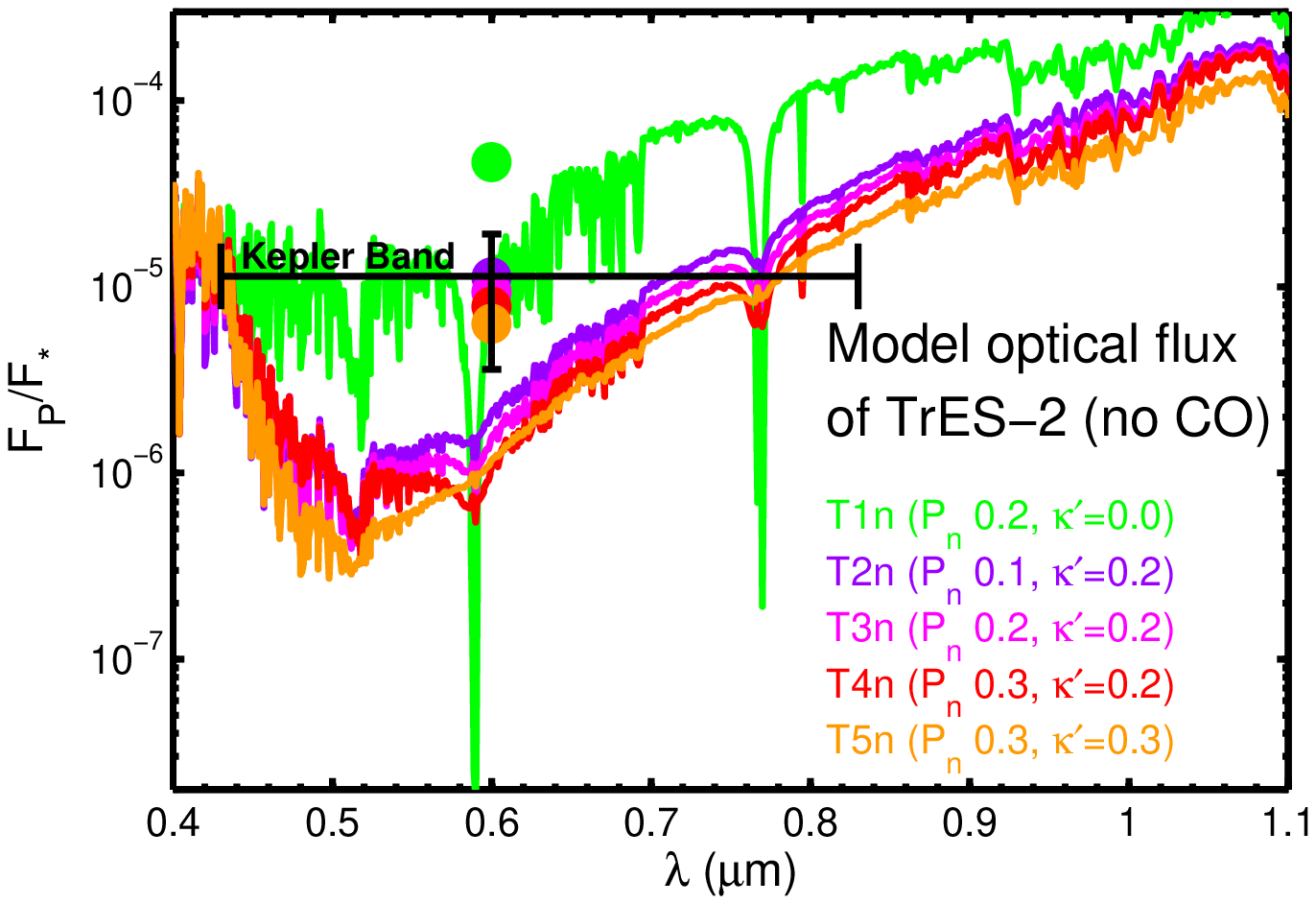}
{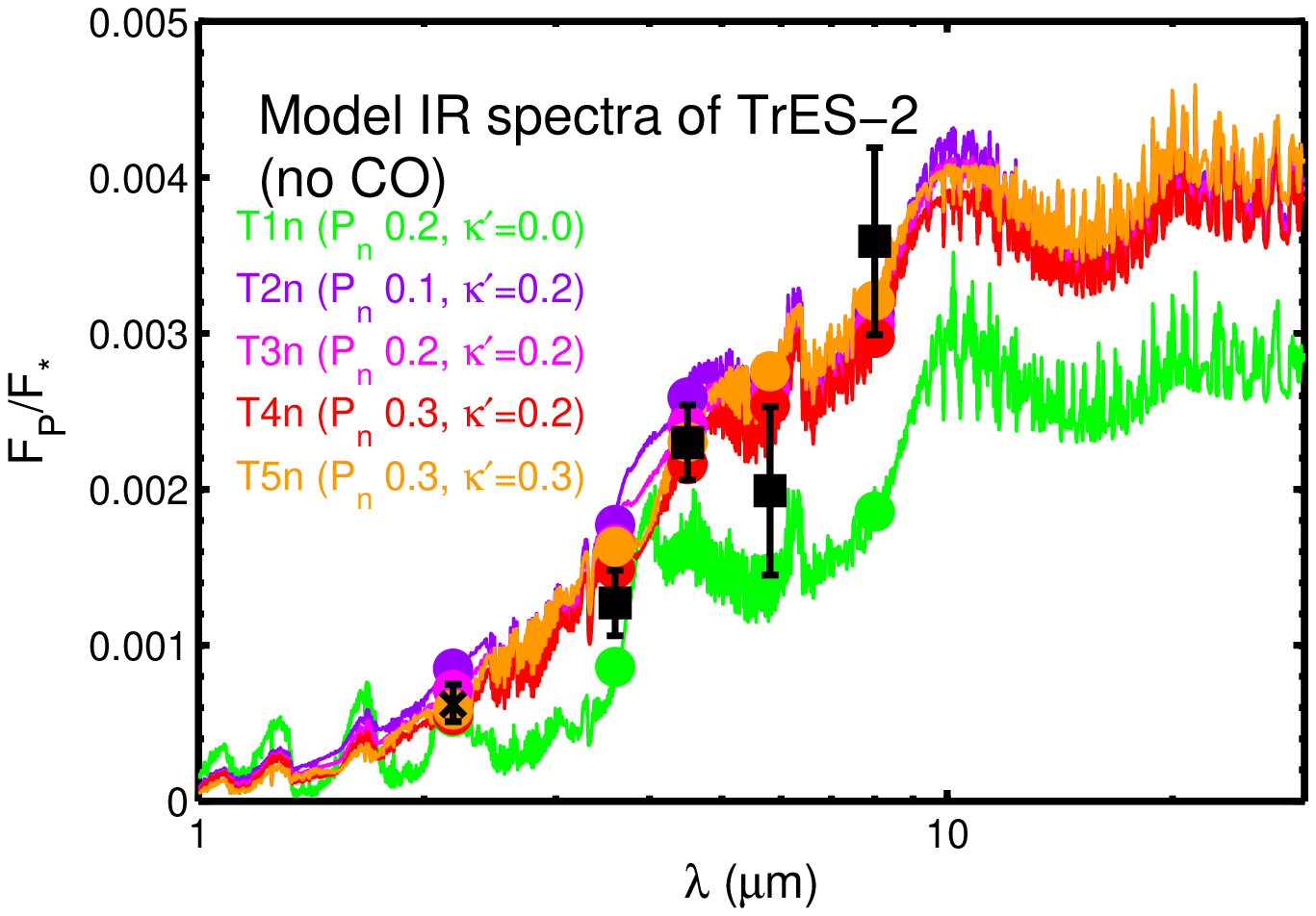}
\plotone
{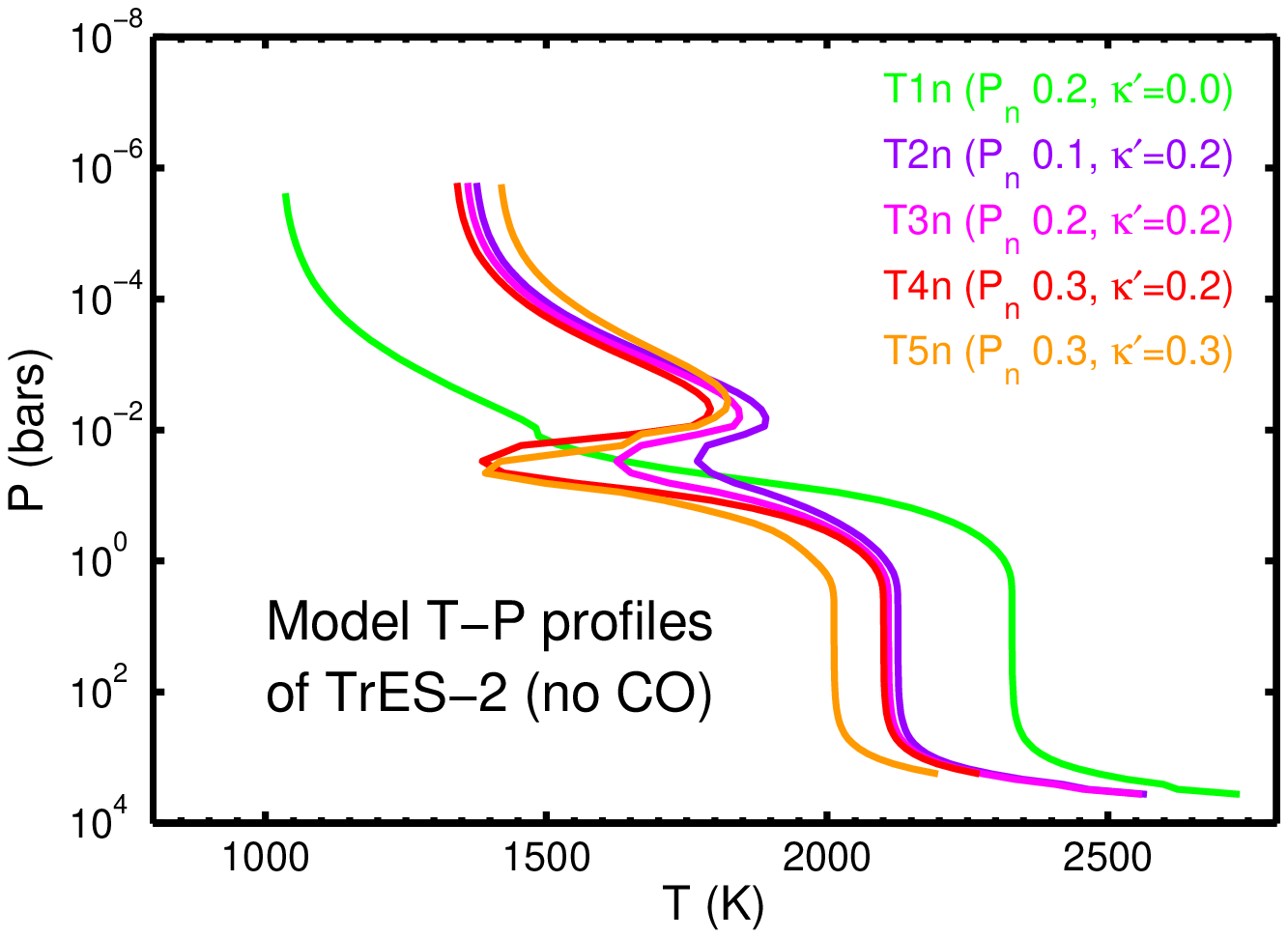}
\caption{TrES-2 nonequilibrium chemistry model atmospheres, analogous
  to Fig.~\ref{fig:T2}, but with CO artificially removed from the
  opacities and with CH$_4$ and H$_2$O correspondingly enhanced.
  Below, we compare with the models shown in Fig.~\ref{fig:T2}. {\it
    Top left:} Optical planet-star flux ratios, with
  \citet{kipping+bakos2010} data (including 2-$\sigma$ error bars)
  shown.  As in Fig.~\ref{fig:T2}, models with upper atmosphere
  absorption (T2n--T5n) tend to have low optical flux ($F_p/F_* \lsim
  2\times 10^{-5}$).  With $\kappa' = 0\rm~cm^2~g^{-1}$ (model T1n),
  the optical flux ratio is $\sim$5$\times 10^{-5}$, higher than the
  corresponding model in Fig.~\ref{fig:T2}, but still low compared
  with the HAT-P-7b data and models.  Models T2n--T5n are all
  consistent with the data, to 2~$\sigma$.  {\it Top Right:} Infrared
  planet-star flux ratios.  Again, the IRAC data clearly favor models
  with some extra absorption.  The 5.8-$\mu$m-discrepancy for model
  T5n is significantly reduced compared with the corresponding model
  with equilibrium chemistry.  With CO removed, model T4n is a
  reasonably good fit to the data (within $\sim$1$\sigma$ at all IRAC
  bands). {\it Bottom:} Temperature-pressure profiles.  The overall
  features of the profiles are similar, but the detailed shapes of the
  profiles are somewhat different, with isothermal regions $\sim$100~K
  warmer and upper atmospheres $\sim$100~K cooler than the equilibrium
  chemistry models.}
\label{fig:T2n}
\end{figure}

Figures~\ref{fig:T2} and \ref{fig:T2n} portray properties of the
atmosphere models with and without CO, respectively.  In these
figures, similar to Fig.~\ref{fig:H7}, the top-left panel shows the
model planet-star flux ratios in the optical and the top-right panel
shows the same in the infrared, while the bottom panel shows the
temperature-pressure profiles.

We consider first the equilibrium chemistry models in
Fig.~\ref{fig:T2}.  The non-inverted model (T1) badly fails to
reproduce the IRAC data at 4.5~$\mu$m and at 8.0~$\mu$m.  The models
with inversions (T2--T5) have similar temperature-pressure profiles to
one another.  The main difference among these models is the
temperature in the region of redistribution
(10$^{-2}$--10$^{-1}$~bars), where the temperature dip ranges between
$\sim$200~K (T2; $P_n=0.1$) and $\sim$700~K (T4, T5; $P_n=0.3$).  As a
consequence of their similarity, these four profiles correspond to
very similar flux ratios at 4.5~$\mu$m, 5.8~$\mu$m, and 8.0~$\mu$m.
Model T5 ($\kappa'= 0.3\rm~cm^2~g^{-1}$) is the best fit at
3.6~$\mu$m, but the inverted models all overpredict the
5.8~$\mu$m-point by $\sim$2$\sigma$.  Models T2--T5 all predict {\it
  Ks}-band flux consistent with the \citet{croll_et_al2010}
observations.  All five models predict very low planet-star flux
ratios in the optical, in contrast to the models and the {\it Kepler}
observation of HAT-P-7b (for which $F_p/F_*=1.3\times10^{-4}$; see
\S\ref{sec:H7}).  Aside from model T1, which is clearly disfavored by
the IRAC data, the other models all predict planet-star flux ratios of
$\lsim$$2\times10^{-5}$.  With no extra absorber, model T1 still
predicts a Kepler-band flux ratio of only $\sim$3$\times 10^{-5}$.
Models T2--T5 are all consistent with the \citet{kipping+bakos2010}
observation of $(1.1\pm 0.7)\times10^{-5}$ (2-$\sigma$ errors).

The models without CO opacity (T1n--T5n), portrayed in
Fig.~\ref{fig:T2n}, are qualitatively quite similar to the equilibrium
models in all our diagnostics (optical flux, infrared flux, and
thermal profile).  However, there are some differences in detail.  The
deep isothermal layers of all five models are hotter by $\sim$150~K
than their equilibrium counterparts, and the upper atmosphere of the
inverted models (T2n--T5n) is cooler by a comparable amount, while the
upper atmosphere of the non-inverted model (T1n) is still warmer than
its equlibrium analog (T1).  The magnitude of the temperature dip in
the redistribution range is also somewhat muted compared with the
equilibrium models.  These slight $T$-$P$ profile differences, in
conjunction with the altered opacities, result in slightly different
optical and infrared spectra.  In the infrared, the inverted no-CO
models produce lower flux at 5.8~$\mu$m, overpredicting the observed
data by less than the equilbrium models.  Among no-CO models, the IRAC
data are best fit by a model with slightly less extra optical
absorption (T4n, $\kappa'=0.2\rm~cm^2~g^{-1}$).  These models have
optical fluxes that are qualitatively similar to their equilibrium
counterparts, all predicting {\it Kepler} band flux ratios of
$\lsim$$5\times 10^{-5}$.  In particular, model T1n predicts slightly
higher optical flux than does model T1, but the inverted models
T2n--T5n predict slightly lower optical fluxes than do models T2--T5.
The predicted {\it Ks}-band and Kepler-band fluxes of models T3n--T5n
are all consistent with observations of \citet{croll_et_al2010} and
\citet{kipping+bakos2010}, respectively.

The basic conclusion that we draw from examining models with reduced
CO opacity is that even such a drastic reduction of CO abundance as
entirely eliminating it does not cause dramatic changes in thermal
profiles or spectra.  In the analysis of \citet{odonovan_et_al2010},
changing the [CO]:[CH$_4$]:[CO$_2$] ratio from $10^{-6}:10^{-6}:0$
(i.e., 1:1:0) to $10^{-4}:5\times 10^{-5}:2\times 10^{-6}$ (i.e.,
50:25:1), together with changing the temperature-pressure profile from
a non-inverted to an inverted one, results in a modest improvement in
the quality of the fit (particularly at 8-$\mu$m).  We note that, in
addition to having a surprisingly nonequilibrium [CO]:[CH$_4$] ratio,
the \citet{odonovan_et_al2010} noninverted model has a very low total
abundance of carbon.  Even though their inverted model is
substantially sub-solar in carbon (by more than an order of
magnitude), reducing the carbon abundance by another two orders of
magnitude is nearly tantamount to removing carbon from the opacities
entirely.  In sum, our analysis shows that, among models with solar
abundance of carbon, the presence or absence of CO in the database
does not make a large difference in the thermal profiles or in the
emergent spectra, and in either case way a hot upper atmosphere and
thermal inversion are required in the model in order to come at all
close to matching the IRAC data.  Individual molecular abundances
substantially different from what one would obtain for solar elemental
abundances might allow for marginally improved fits to the data, but
might not yet be called for by the relatively sparse data available so
far.

The upper left panels of both Figs.~\ref{fig:T2} and \ref{fig:T2n}
both show that models with extra optical absorbers in the upper
atmosphere have {\it lower} planet-star flux ratios in the {\it
  Kepler} band than the models without an extra absorber.  We revisit
this point in \S\ref{sec:nonmon}.  In light of this generic trend, and
since among radiative equilbrium models the IRAC data are better fit
by inverted models than by non-inverted models, we predict that {\it
  Kepler} photometry of TrES-2 will indicate low optical flux from
this planet, not more than one or a few times $10^{-5}$ of the stellar
flux.  If future {\it Kepler}-band observations reveal optical flux in
excess of this amount, that might be indicative of extra optical
scattering opacity that was not included in our models.

\section{Optical Flux vs. Optical Absorber Strength}
\label{sec:nonmon}
Here, we point out a puzzle: the upper left panel of Fig.~\ref{fig:H7}
shows a different trend of optical flux vs. $\kappa'$ from the
analogous panels of Figs.~\ref{fig:T2} and \ref{fig:T2n}.
\citet{lopez-morales+seager2007} noted that inverted models, with
their hot upper atmospheres, might be expected to have higher optical
flux than non-inverted models.  But how do we explain the trend seen
for TrES-2 (Figs.~\ref{fig:T2} and \ref{fig:T2n})?

If there is an extra absorber in the optical part of the spectrum (a
$\kappa_e$, in the terminology of \citealt{burrows_et_al2008}, or a
$\kappa'$ in the present work), the emergent optical flux is affected
by two competing effects.  The absorber makes the planet darker at
altitude, but can also heat the upper atmosphere.  The latter effect
makes the upper atmosphere of the planet more emissive.

\begin{figure}
\plotone{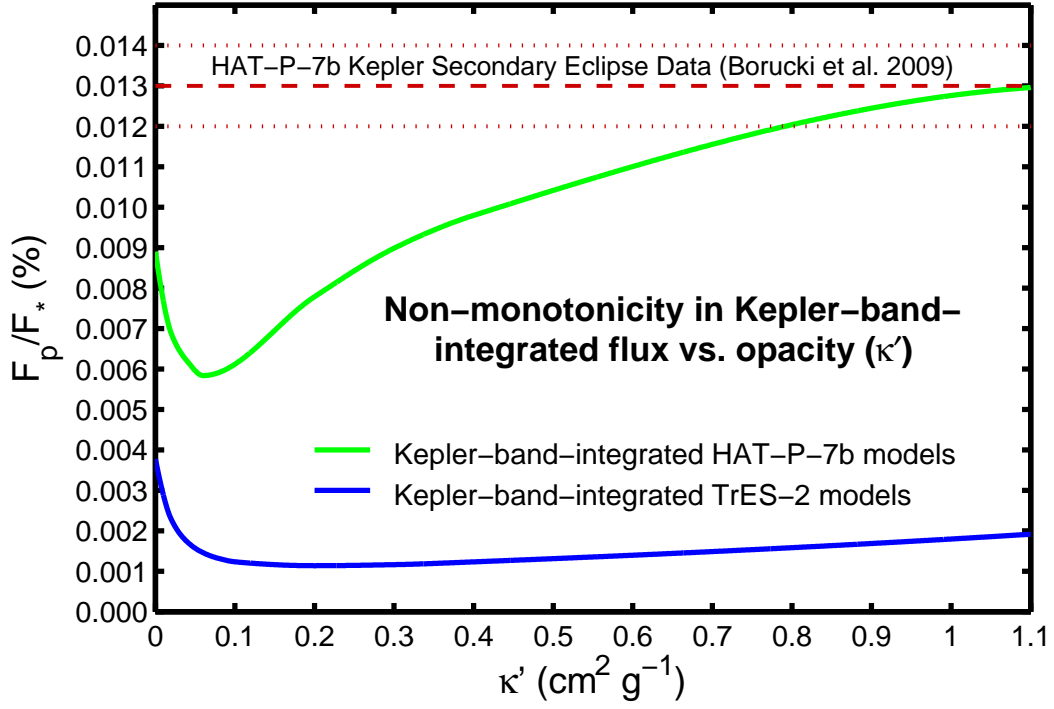}
\caption{{\it Kepler} bandpass brightness ratios vs. upper atmosphere
  absorption ($\kappa'$).  Models of HAT-P-7b (thick light green) and
  TrES-2 (thick blue) are shown, in addition to the {\it Kepler} data
  for HAT-P-7b (dashed red) with 1-$\sigma$ errors (dotted dark
  green).  These models have no redistribution to the nightside ($P_n
  = 0$).  The relationship between optical flux and $\kappa'$ is
  non-monotonic for both planets, but more so for HAT-P-7b.  In
  particular, for HAT-P-7b, large enough values of $\kappa'$ (values
  above $\sim$0.3~cm$^2$~g$^{-1}$) lead to enough flux in the region
  of the Wien tail that the optical flux is greater than it would be
  with no extra absorber.  However, for TrES-2, despite the
  non-monotonicity, the planet is still brightest in the optical
  without any extra absorber.}
\label{fig:nonmon}
\end{figure}

Figure~\ref{fig:nonmon} illustrates how the balance of these two
effects depends on the opacity ($\kappa'$) of the upper-atmosphere
absorber, in both HAT-P-7b and TrES-2.  The {\it Kepler} bandpass
brightness of $P_n=0.0$ models of both planets (HAT-P-7b: green;
TrES-2: blue) is plotted as a function of $\kappa'$, for a series of
values of $\kappa'$ between 0 and 1.1$\rm~cm^2~g^{-1}$.  For both
planets, a small amount of extra absorption results in reduced
emergent flux in the {\it Kepler} band, and, for both planets, large
values of $\kappa'$ ($\gsim$0.3~cm$^2$~g$^{-1}$) result in increased
optical emission, as emission from the Wien tail of the hot upper
atmosphere becomes more prominent in the optical part of the spectrum.
There is, therefore, a generic non-monotonic character to the
dependence of optical flux on $\kappa'$.  The degree of the
non-monotonicity, however, is greater for HAT-P-7b models than for
TrES-2 models.  In the former, the optical flux for $\kappa' \gsim
0.3\rm~cm^2~g^{-1}$ is greater than with no extra absorption, while in
the latter, even with $\kappa'=1.1\rm~cm^2~g^{-1}$ the optical flux is
just over half of what it is with no extra absorption.  The origin of
this difference is related to the properties of the irradiation.
HAT-P-7 is an exceptionally large ($1.84R_\sun$) star that is fairly
hot (F6, 6350~K), whereas TrES-2's star is a more ordinary solar-type
star ($1.003R_\sun$, G0V, 5850~K).  As a result, the incident optical
irradiation at HAT-P-7b is more than 4 times as great as on TrES-2.
This greatly enhanced incident optical irradiation results in a much
greater sensitivity to the presence of an extra optical absorber.

\section{Conclusion}
\label{sec:conc}
We have presented atmosphere models of HAT-P-7b and TrES-2, two of the
three {\it Kepler} field planets that were known prior to the start of
the {\it Kepler} mission.  We find that the combination of the IRAC
and {\it Kepler} secondary eclipse data for HAT-P-7b, with a {\it
  Kepler}-bandpass secondary eclipse ratio of $\sim$$1.3\times
10^{-4}$, appear to require an extremely hot upper atmosphere, with an
extra optical absorber that creates a strong thermal inversion and
with little day-night redistribution.  The IRAC data for TrES-2 led us
us to expect that TrES-2 has a much lower planet-star flux ratio
($\lsim$$2\times 10^{-5}$) in the {\it Kepler} bandpass than does the
HAT-P-7 system, and indeed this is what was seen in recently published
{\it Kepler} data of this object.

Furthermore, we find that there is a non-monotonic relationship
between $\kappa'$ and a planet's day-side emergent optical flux.  This
non-monotonicity highlights the need for multiwavelength observations
in order to better estimate the atmospheric structure.

\vspace{0.3in}

\acknowledgments

We thank Ivan Hubeny, Laurent Ibgui, Kevin Heng, and Jason Nordhaus
for useful discussions.  We also appreciate the careful reading and
helpful comments from the referee, Giovanna Tinetti.  This study was
supported in part by NASA grant NNX07AG80G.  We also acknowledge
support through JPL/Spitzer Agreements 1328092, 1348668, and 1312647.
The authors are pleased to acknowledge that part of this research was
performed while in residence at the Kavli Institute for Theoretical
Physics, and was supported in part by the National Science Foundation
under Grant No. PHY05-51164.

\vspace{1.2in}

\begin{appendix}
\label{sec:app}
\section*{Parameterizing Redistribution in a 1D Model}
\label{ssec:redist}
There have been several approaches used in the literature for treating
the redistribution of day-side irradiation to the nightside in the
context of one-dimensional models.  Here, we consider several
parameterizations and the relationships among them.  Early work,
including that of our group prior to \citet{burrows_et_al2006}, used
only a single parameter (called ``$f$'') to describe redistribution.
However, there are at least three effects that ought to be included in
a description of how an atmosphere redistributes heat: (1) some
fraction of incident energy is transported to the nightside via
atmospheric motions (we call this fraction ``$P_n$''); (2) this
redistribution occurs at some depth between the top and the bottom of
the atmosphere; and (3) the visible face of the planet (the dayside at
secondary eclipse phase) has (in general) an anisotropic distribution
of specific intensity in the direction of Earth.  Our more recent
work, including this paper, employs our attempt, however imperfect, to
incorporate these physical effects that 1D models from other groups
have not fully included.  In particular, when other modeling efforts
have implemented schemes for day-night heat transport, they have
essentially taken the redistribution to occur before the incoming
radiation reaches the top of the atmosphere.  In contrast, we specify
the range of pressures at which the redistribution occurs; while our
numerical choice might be not be accurate in detail, it is a
physically motivated, less ad hoc way to parameterize the physics that
we know affect emergent spectra.

One of the best-known redistribution parameters is the afforementioned
geometrical $f$ factor, which arises from an energy balance relation
of the following form:
\begin{eqnarray}
\label{eq:f} L_p & = & \frac{\pi}{f} R_p^2 \sigma T_{\rm unif}^4 \, , \\
\label{eq:f2} \mbox{or,} \quad \frac{F_p}{F_*} & = & f \left( \frac{R_p}{a} \right)^2 \, .
\end{eqnarray}
In eq.~(\ref{eq:f}), $L_p \equiv (\pi R_p^2) L_*/ (4 \pi a^2)$ is the
total stellar power intercepted by the planet (and, therefore,
approximately equals the emergent radiation from the planet), $R_p$ is
the planet's radius, $L_*$ is the stellar luminosity, $a$ is the
orbital separation, $\sigma$ is the Stefan-Boltzmann constant, and
$T_{\rm unif}$ is the temperature of a uniform-temperature sphere of
radius $R_p$ whose flux at Earth near secondary eclipse phase would be
the same as that of the planet.  In eq.~(\ref{eq:f2}), $F_p$ is the
integrated flux from the planet at Earth,\footnote{$F_p = \sigma
  T_{\rm unif}^4 (R_p/d)^2$, where $d$ is the distance from the planet
  to Earth.} and $F_*$ is the integrated stellar flux.  In
eqs.~(\ref{eq:f}) and (\ref{eq:f2}), and in the remainder of this
section, we have ignored both scattering in the planet's atmosphere
and its own intrinsic luminosity (from its heat of formation, tidal
heating, etc.).  The $f$ factor has been used by a number of authors
in the last decade (\citealt{burrows_et_al2003, burrows_et_al2005,
  fortney_et_al2005, lopez-morales+seager2007, hansen2008}; and
referred to as $\alpha$ in \citealt{barman_et_al2005}).\footnote{We
  note that \citet{burrows_et_al2008b} also use a parameter called
  $f$, but this $f$ (which is typically set to 2/3) represents the
  direction-cosine of incident irradiation that is used in the planar
  atmosphere calculation, and is therefore related to distribution of
  temperature on the dayside, instead of to the fraction of energy
  that is redistributed to the nightside.  Some work (e.g. that of
  \citealt{barman_et_al2005}) integrates the contributions to the
  total planet-flux at Earth of a series of concentric annuli, from
  the substellar point to the terminator.  \citet{burrows_et_al2008b}
  show that this integral is extremely well-approximated by taking the
  entire visible hemisphere as being irradiated by the ray at
  direction-cosine 2/3.}  Perfect redistribution (implying that the
planet is itself of uniform temperature) corresponds to $f = 1/4$.
Zero redistribution, so that each annulus a given angle away from the
substellar point absorbs and reradiates its local irradiation
isotropically, corresponds to $f=2/3$.\footnote{In this case, the
  specific intensity in the direction of Earth from a annulus at
  direction-cosine $\mu$ away from the substellar point (at full-moon
  phase) is proportional to $\mu$.  Specifically, $I[\mu] = \mu L_* /
  (2 \pi a)^2$.}  There is, thus, a ``beaming factor'' that increases
$f$ by a factor of 4/3 over the value (1/2) that it would have if the
dayside were of uniform temperature instead of peaked toward the
center of the disk in accordance with the irradiating flux.  $f$ is
typically implemented in 1D atmosphere models as simply a uniform
reduction of the incident flux at the top of the atmosphere
(algorithmically, though not physically, equivalent to the effect of
albedo).

Alternatively, we may characterize the redistribution by removing a
fraction $P_n$ of the incident stellar flux on the dayside in a
prescribed pressure interval (as described in \S\ref{sec:methods}),
and inserting the energy at a similar (or different) level on the
nightside.  In such a prescription, $P_n$ plausibly ranges between 0
(corresponding to no redistribution) and 0.5 (corresponding to the
nightside radiating as much power as the dayside, via redistributive
winds).  \citet{burrows_et_al2006}, \citet{burrows_et_al2008b}, and
other work from our group implement this procedure.  In order to
calculate theoretical secondary eclipse fluxes and spectra, one must
have a model of the three-dimensional distribution of temperature in a
planet's atmosphere (e.g., \citealt{showman_et_al2008,
  showman_et_al2009, burrows_et_al2010}).  In lieu of performing
three-dimensional dynamical calculations, our 1D atmosphere models
assume the same beaming factor for day-side emergent radiation,
regardless of the total day-side radiance; and they assume uniform
temperature on the nightside.  In particular, on the dayside, when
$P_n$ = 0.5, $f=1/3$ (which is 4/3 of the 1/4 that $f$ would equal for
a uniform-temperature dayside, when half the incident flux has been
redistributed to the nightside).  More generally, the integrated model
day-side flux from the planet can be approximately described by
eq.~(\ref{eq:f2}), taking $f$ to be defined by
\begin{equation}
f = \frac{2}{3}(1-P_n) \, .
\label{eq:fPn}
\end{equation}
The descriptions in \citet{cowan_et_al2007} of how ``$T_{\rm day}$''
and ``$T_{\rm night}$'' depend on $P_n$ correctly quantify the total
emission from the day and night sides of the planet, but they do not
include the beaming caused by an anisotropic temperature distribution
on the dayside that is peaked at the substellar point; therefore,
unlike what is suggested in \citet{cowan_et_al2007}, the integrated
day-side flux is not described by taking the dayside to be of uniform
temperature $T_{\rm day}$.

Yet a third redistribution parameter, $\varepsilon$, is suggested by
\citet{cowan+agol2010}.  $\varepsilon$ ranges between 0 (no
redistribution) and 1 (full redistribution).  It initially appears to
be defined similarly to $P_n$, and is described (in part) by the
following relation:
\begin{equation}
\label{eq:epsday} L_p = \frac{\pi}{(8 - 5\varepsilon)/12} R_p^2 \sigma \widetilde{T}_{\rm day}^4 \, ,
\end{equation}
where $\widetilde{T}_{\rm day}$ is the same as $T_{\rm unif}$ of
eq.~(\ref{eq:f}), i.e., a measure of the flux in the direction of
Earth.  In eq.~(\ref{eq:epsday}), there is a tilde above the
expression for day-side temperature to distinguish it from the
corresponding expression in \citet{cowan_et_al2007}, which is a
measure of the total day-side emission, as opposed to the flux in the
direction of Earth.  Comparing eqs.~(\ref{eq:f}) and (\ref{eq:epsday})
shows that $\varepsilon$ is a rescaling of $f$: $\varepsilon = (8 - 12
f)/5$.  \citeauthor{cowan+agol2010} suggest that their $\varepsilon=0$
limit produces a brighter dayside than does the $P_n=0$ limit, but it
does not.  Instead, the $P_n=0.5$ limit produces a brighter dayside
than does the $\varepsilon=1$ limit (brighter by the beaming factor of
4/3).

Finally, we emphasize that while eqs.~ (\ref{eq:f2}) and
(\ref{eq:fPn}) describe how the integrated planet flux depends on
$P_n$, this parameter does more than simply affect the
top-of-atmosphere energy budget.  The real benefit of $P_n$ (which is
not captured by the above equations) is that, when nonzero, instead of
reducing incident irradiation, it removes energy from the dayside (and
deposits it on the nightside) at more realistic levels in the
atmosphere.  This process contributes to a slight thermal inversion by
cooling the middle atmosphere, and, therefore, affects the ratio of
3.6-$\mu$m flux to the 4.5-$\mu$m flux in a way that is not reproduced
by the standard $f$ parameterizations in which incident irradiation is
reduced at the top of the atmopshere.

\end{appendix}

\newpage

\bibliography{biblio.bib}

\newpage

\begin{table}[ht]
\small
\begin{center}
\caption{HAT-P-7\MakeTextLowercase{b} and TrES-2 models} \label{ta:models}
\begin{tabular}{l|ccl}
\hline
\hline
                    &         &                    &                 \\[0.2cm]
\multirow{2}{*} {Model} & {$P_n$} & {$\kappa'$}         & \multirow{2}{*} {Chemistry} \\
                        & {}      & {(cm$^2$~g$^{-1}$)} &                              \\
\hline
\rule {-3pt} {10pt}
H1\tablenotemark{a} & 0.10    &  0.0               &  equilibrium    \\[0.2cm]
H2                  & 0.05    &  0.4               &  equilibrium    \\[0.2cm]
H3                  & 0.10    &  0.7               &  equilibrium    \\[0.2cm]
H4                  & 0.10    &  1.1               &  equilibrium    \\[0.2cm]
H5                  & 0.00    &  1.1               &  equilibrium    \\[0.2cm]
  \tableline
                    &         &                    &                 \\[0.2cm]
T1\tablenotemark{b} & 0.2     &  0.0               &  equilibrium    \\[0.2cm]
T2                  & 0.1     &  0.2               &  equilibrium    \\[0.2cm]
T3                  & 0.2     &  0.2               &  equilibrium    \\[0.2cm]
T4                  & 0.3     &  0.2               &  equilibrium    \\[0.2cm]
T5                  & 0.3     &  0.3               &  equilibrium    \\[0.2cm]
  \tableline
                    &         &                    &                 \\[0.2cm]
T1n                 & 0.2     &  0.0               &  no CO\tablenotemark{c} \\[0.2cm]
T2n                 & 0.1     &  0.2               &  no CO    \\[0.2cm]
T3n                 & 0.2     &  0.2               &  no CO    \\[0.2cm]
T4n                 & 0.3     &  0.2               &  no CO    \\[0.2cm]
T5n                 & 0.3     &  0.3               &  no CO    \\[0.2cm]
\end{tabular}
\tablenotetext{a}{Models H1--H5 have the following parameters --
  orbital semimajor axis ($a$), planet and stellar surface gravity
  ($g_p$, $g_*$), planet and stellar radii ($R_p$, $R_*$), and stellar
  effective temperature $T_*$ -- appropriate to the HAT-P-7 system ($a
  = 0.0377$~AU; $\log_{10}g_p = 3.38$, $\log_{10}g_* = 4.08$ where
  $g_p$, $g_*$ are in cgs; $R_p = 1.363 R_J$, where $R_J \equiv
  7.15\times 10^9$~cm is Jupiter's radius; $R_* = 1.84 R_\sun$; $T_* =
  6350$~K).  The irradiating spectrum is interpolated from Kurucz
  models.\\}
\vspace{0.1in}
\tablenotetext{b}{Models T1--T5 and T1n--T5n have parameters
  appropriate to the TrES-2 system ($a = 0.0356$~AU; $\log_{10}g_p =
  3.30$, $\log_{10}g_* = 4.43$ where $g_p$, $g_*$ are in cgs; $R_p =
  1.224 R_J$; $R_* = 1.003 R_\sun$; $T_* = 5850$~K).\\}
\vspace{0.1in}
\tablenotetext{c}{Models T1n--T5n have opacities defined by
  out-of-equilibrium chemistry, where CO has been artificially
  removed, with the carbon instead in CH$_4$ and the excess oxygen
  instead in H$_2$O.}
\end{center}
\end{table}

\clearpage


\clearpage


\clearpage

\clearpage

\end{document}